\documentclass[trackchanges,preprint2]{aastex62}

\usepackage{amsmath}
\usepackage{breqn}
\usepackage{verbatim}
\usepackage{multirow}
\usepackage{booktabs}
\usepackage{bm}

\submitjournal{ApJL}

\shorttitle{Recent Massive Black Hole Merger in the Galactic Center}
\shortauthors{Akiba, Naoz, \& Madigan}

\begin{document}

\title{On the Formation of S-stars from a Recent Massive Black Hole Merger in the Galactic Center}

\author[0000-0002-0647-718X]{Tatsuya Akiba}
\affiliation{JILA and Department of Astrophysical and Planetary Sciences, CU Boulder, Boulder, CO 80309, USA}
\email{tatsuya.akiba@colorado.edu}

\author[0000-0002-9802-9279]{Smadar Naoz}
\affiliation{Department of Physics and Astronomy, University of California, Los Angeles, CA 90095, USA}
\affiliation{Mani L. Bhaumik Institute for Theoretical Physics, University of California, Los Angeles, Los Angeles, CA 90095, USA }

\author[0000-0002-1119-5769]{Ann-Marie Madigan}
\affiliation{JILA and Department of Astrophysical and Planetary Sciences, CU Boulder, Boulder, CO 80309, USA}

\begin{abstract}

The Galactic Center hosts a rotating disk of young stars between 0.05 and 0.5 pc of Sgr A*. The “S-stars” at a distance $<0.04$ pc, however, are on eccentric orbits with nearly isotropically distributed inclinations. The dynamical origin of the S-star cluster has remained a theoretical challenge. Using a series of $N$-body simulations, we show that a recent massive black hole merger with Sgr A* can self-consistently produce many of the orbital properties of the Galactic nuclear star cluster within 0.5 pc. A black hole merger results in a gravitational wave recoil kick, which causes the surrounding cluster to form an apse-aligned, eccentric disk. We show that stars near the inner edge of an eccentric disk migrate inward and are driven to high eccentricities and inclinations due to secular torques similar to the eccentric Kozai-Lidov mechanism. In our fiducial model, starting with a thin eccentric disk with $e = 0.3$, the initially unoccupied region within $0.04$ pc is populated with high eccentricity, high inclination S-stars within a few Myr. This dynamical channel would suggest that a black hole of mass $2^{+3}_{-1.2} \times 10^5 \ M_{\odot}$ merged with Sgr A* within the last 10 Myr.

\end{abstract}

\keywords{Galactic center \-- Stellar dynamics \--  Supermassive black holes}

\section{Introduction} \label{sec:intro}

Most massive galaxies, including our own Milky Way, harbor supermassive black holes (SMBHs) in their centers. SMBHs are intimately linked to their host galaxies and are fundamental to our understanding of galaxy evolution \citep[e.g.,][]{Fer00, Springel2005b, dimatteo2005}. The SMBH in our Galaxy, Sgr A*, and the morphology of its surrounding environment are influenced by the region's history. Specifically, in the innermost regions of the nuclear star cluster surrounding Sgr A* is a population of young stars. Between 0.05 and 0.5 pc from the center is a well-defined, clockwise disk of O, B, and Wolf-Rayet stars \citep[e.g.,][]{genzel1997, genzel2000, ghez1998, yelda2014, ghez2000, paumard2006, vonFellenberg2022}. The disk appears to be relatively thin with a thickness of around $10^\circ$ \citep{lu2009, bartko2009}, and exhibits a unimodal eccentricity distribution that peaks at $e \approx 0.3$ \citep[e.g.,][]{yelda2014}, although, see \citet{Naoz+18} for an alternative explanation. Even closer in, within $0.04$ pc, is a nearly isotropic cluster of B stars known as the ``S-stars'' which are much more eccentric, consistent with a thermal distribution in the range $0.3 \leq e \leq 0.95$ \citep{gillessen2017, ali2020}.

The disk stars could have reasonably come from a single star formation episode, for instance, from the fragmentation of an accretion disk \citep[e.g.,][]{levin2003, nayakshin2005a, nayakshin2005b, Lu13}. However, the origin of the S-stars has been a theoretical challenge. The gravitational influence of Sgr A* prevents stars from forming in place at distances less than 0.04 pc \citep{levin2007}. At the same time, dynamical mechanisms for migration must work efficiently based on the young age of S-stars estimated to $\leq 15$ Myr \citep{habibi2017}. A couple of possibilities that have been explored are the tidal disruption of field binaries via the Hills mechanism \citep[e.g.,][]{hills1988, ginsburg2006} and the direct migration of disk stars \citep[e.g.,][]{levin2007, griv2010, chen2014}, but both mechanisms struggle to reproduce the eccentricity and inclination distributions of the observed S-stars. An additional relaxation process must be invoked in either case to thermalize and randomize the eccentricity and inclination distributions, which is expected to take longer than the estimated age of the cluster \citep[e.g.,][]{perets2009, antonini2013, zhang2013}.

\begin{figure*}[tb!]
\centering
\includegraphics[width=\linewidth]{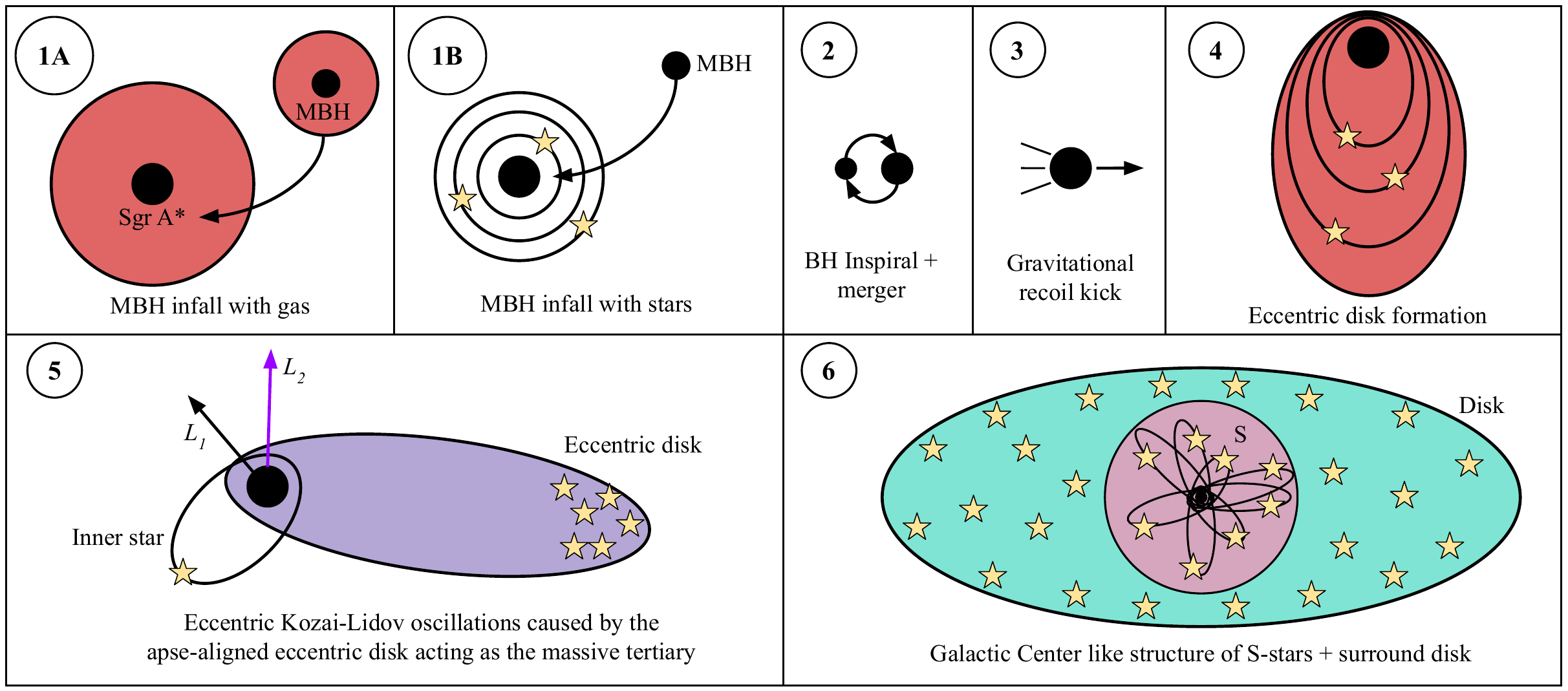}
\caption{A schematic diagram illustrating the proposed process for forming the Galactic Center structure within the central 0.5 pc. 1) A MBH falls toward Sgr A* with accompanying A) gas, B) young stars or a combination of both. 2) The black holes inspiral and merge. 3) The merger remnant receives a gravitational recoil kick. 4) The kick induces an eccentric, apse-aligned disk of stars or gas (from which stars can form). 5) Stars near the inner edge of the eccentric disk undergo eccentric Kozai-Lidov oscillations caused by the asymmetric gravitational potential of the eccentric disk. The inner stars are driven to high eccentricities and inclinations. 6) The S-stars and the surrounding disk are produced within a few Myr of the merger.}
\label{fig:schematic}
\end{figure*}

Here, we propose that a relatively recent merger between two massive black holes (MBHs) at the center of our Galaxy can naturally explain the aforementioned morphology. The formation and evolution of galaxies likely involve the accretion of surrounding globular clusters or dwarf galaxies \citep{volonteri2003, rodriguez-gomez2015, paudel2018, pillepich2018, Hochart+24}, which can provide an avenue for an intermediate-mass black hole (IMBH) or a low-mass SMBH to sink toward the Galactic Center \citep{madau2001, volonteri2003b, rashkov2014}. The presence of an intermediate-mass companion has been invoked to explain the peculiar orbital distributions of the S-star cluster and the surrounding disk \citep[e.g.,][]{merritt2009, zheng2020, Naoz+20, Zhang+23, Will+23, GRAVITY+20, Fouvry+23, ginat2023}, and there are several candidate IMBHs in the vicinity of Sgr A* (see \citealt{greene2020} for a review). Recently, this has become a more intriguing possibility as new high-spin measurements of Sgr A* \citep{eht2023, daly2024} imply that a substantial fraction of its mass may have been accreted to spin up the black hole; \citet{wang2024} showed that a past MBH merger in the Galactic Center can consistently produce the spin of Sgr A*.

During the merger of two MBHs, the anisotropic emission of gravitational waves causes a recoil kick to be imparted on the merger remnant \citep[e.g.,][]{Bekenstein1973, Wiseman92, Herrmann07}. Gravitational recoil kicks of 100 km/s or less are small perturbations and the SMBH returns to the center of the galaxy on $\sim$Myr timescales \citep{Blecha}. Even small kicks can, however, align the eccentricity vectors of stellar orbits surrounding the merged black hole, forming an eccentric stellar disk \citep{Akiba2021, Akiba2023}. An eccentric disk origin for the Galactic Center has been studied in the past \citep[e.g.,][]{Madigan2009, Subr2016, generozov2020, Generozov2022, rantala2024}; the novelty here is in attributing the presence of an eccentric disk to a MBH merger with Sgr A*. We study the dynamical evolution of an eccentric disk to show that many of the orbital properties of the S-star cluster and the surrounding disk are self-consistently produced within a few Myr. We use this dynamical channel to constrain the mass of the companion that may have merged with Sgr A* in the recent past.

The paper is organized as follows. In Section \ref{sec:merger}, we introduce the specific merger scenario we consider in this work. In Section \ref{sec:numerical}, we describe our numerical setup and modeling of relevant physical processes. In Section \ref{sec:proof}, we present key results of our fiducial model, and we conclude by discussing further implications for the Galactic Center in Section \ref{sec:discussion}.

\section{The Merger Scenario}
\label{sec:merger}

\begin{figure}[t!]
\centering
\includegraphics[width=\linewidth]{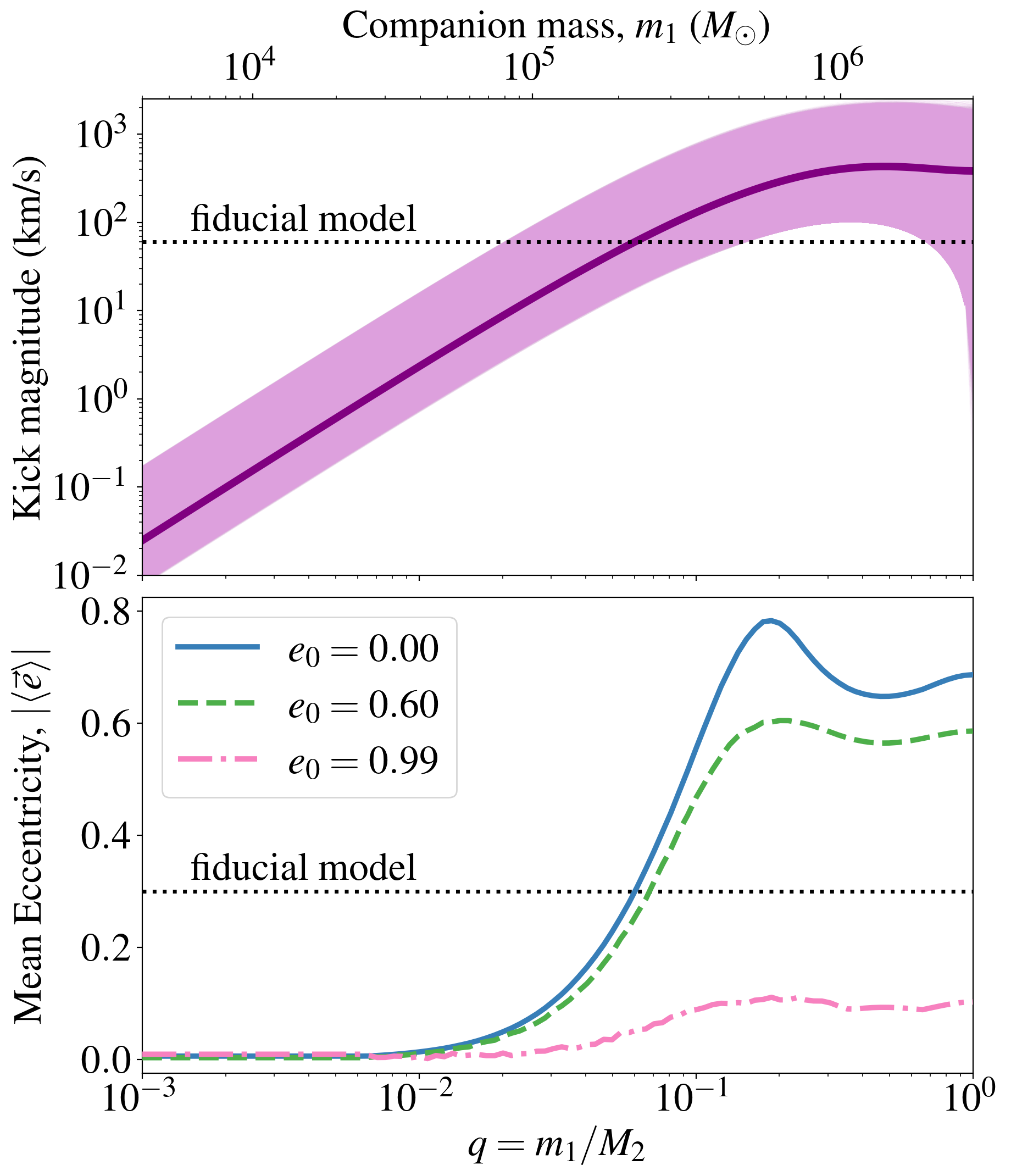}
\caption{(\textit{Top:}) Kick magnitude as a function of the mass ratio, $q$, according to the analytical prescription of \citet{lousto2010, lousto2012}. The light purple region shows $10^6$ Monte Carlo evaluations assuming a random spin distribution where the dimensionless spin parameter is uniform between 0 and 1 and isotropic in direction for both of the pre-merger black holes. The thick, purple line shows the average kick magnitude-dependence obtained from the $10^6$ trials. (\textit{Bottom:}) Alignment of the eccentricity vectors, $| \langle \vec{e} \rangle |$, in the range $a = 0.04$--$0.5$ pc as a function of the mass ratio, $q$, of the black holes assuming the average result from the top panel. $e_0$ corresponds to the pre-kick eccentricities in the stellar disk. $m_1$ and $M_2$ are the masses of the less massive and more massive black hole, respectively. The dotted black line corresponds to our fiducial model run of $e = 0.3$ in both panels.}
\label{fig:bh_kick}
\end{figure}

The merger scenario we consider in this work is schematically illustrated in Figure \ref{fig:schematic}. (1) A MBH falls toward Sgr A* in the presence of (A) gas or (B) stars. Initially, we assume an axisymmetric distribution of (gas or stellar) orbits surrounding Sgr A*. (2) As the massive companion inspirals and merges with Sgr A*, (3) a gravitational recoil kick is imparted on the merger remnant, and (4) causes the surrounding disk to become eccentric and apsidally-aligned. If this is an eccentric gas disk, we assume that a star formation episode can be triggered following the merger process so that we have an eccentric stellar disk around the recoiling black hole. (5) We run a series of numerical experiments to show that eccentric disks undergo a dynamical process similar to the eccentric Kozai-Lidov (EKL) mechanism, and (6) produce many of the orbital properties of the S-stars and the clockwise disk within a few Myr. Specifically, we seek to reproduce the following orbital properties of the Galactic Center:
\begin{enumerate}
    \item Within $0.04$ pc, stars have a nearly isotropic distribution of inclinations and a thermal eccentricity distribution, $N(e)=2e \ de$, in the range $0.3 \leq e \leq 0.95$.
    \item Between 0.05 and 0.5 pc, stars are in a coherently-rotating disk with a thickness of around $10^\circ$ and mean $e \approx 0.3$.
\end{enumerate}

First, we explore the eccentric disk formation process. In \citet{Akiba2021, Akiba2023}, the formation of an eccentric disk via the recoil kick was explored in detail, but the relationship between the kick and the properties of the pre-merger black holes was not considered. Here, we include the mapping between the mass ratio of the black holes and the expected kick magnitude to study the degree of post-kick apsidal alignment we expect, given the mass of the companion.

To obtain the kick magnitude, we use the analytical model from \citet{lousto2010, lousto2012} as outlined in \citet{fragione2023}. Their prescription takes into account the three-dimensional spin vectors of both black holes (see Appendix \ref{app:kick} for details) and has shown good agreement with full numerical relativity results out to the moderately unequal mass ratio regime of $q \sim 0.1$ \citep{gonzalez2009}. The predicted kick magnitude as a function of the mass ratio, $q$, is shown in Figure \ref{fig:bh_kick} on the top panel. The light purple region shows the Monte Carlo results of $10^6$ evaluations of the analytical model with randomized spins where the dimensionless spin angular momenta of both black holes are drawn from an isotropic distribution in direction and a uniform distribution between 0 and 1 in magnitude. The thick purple line shows the average kick magnitude as a function of $q$ marginalized over the assumed spin distribution.

Kick magnitudes upward of 100 km/s are expected for mass ratios $q > 0.1$ and rapidly decay as one moves toward more extreme mass ratios. The average shows a peak of around 460 km/s at a mass ratio of $q \approx 0.5$. Using this average kick magnitude-dependence on $q$, we run a suite of \texttt{REBOUND} \citep{Rein2012} simulations of the surrounding nuclear star cluster to study the instantaneous alignment of stellar eccentricity vectors in the semi-major axis range 0.04--0.5 pc, consistent with the current clockwise disk. We initialize $N = 5 \times 10^4$ stars in a thin disk where the inclination is Rayleigh-distributed with scale parameter $\sigma_i = 3^\circ$ and every star begins at an initial eccentricity, $e_0$, prior to the kick. An impulsive in-plane kick is applied to the central black hole with mass $M = 4 \times 10^6 \ M_{\odot}$, and we record the instantaneous change in each of the stellar orbits. The pre-kick circular disk is, to first-order, motivated by \citet{mb2023} who found tangentially anisotropic stellar distributions following a MBH binary inspiral. We choose to simulate an in-plane kick because they are statistically more likely even if the kick direction was isotropically distributed \citep{akiba2024}, and gas torques have been shown to cause alignment of the black hole orbit and spins skewing the distribution more toward in-plane kicks \citep{Bogdanovic2007}.

We quantify apsidal alignment using the mean eccentricity vector,
\begin{equation}
\langle \vec{e} \rangle = \frac{\sum_{i} \vec{e}_i}{N} \ ,
\end{equation}
where $\vec{e}_i$ is the eccentricity vector of the $i$-th particle, and $N$ is the number of stars. The magnitude of this vector is a measure of apsidal alignment where 0 indicates no alignment and $| \langle \vec{e} \rangle | = \langle e \rangle$ (the average scalar eccentricity) when eccentric orbits are perfectly aligned. We note that this measure is sensitive to both the scalar eccentricity of orbits as well as their degree of apsidal alignment.

In Figure \ref{fig:bh_kick} on the bottom panel, we show the magnitude of the mean eccentricity vector, $| \langle 
\vec{e} \rangle |$, induced in the $a = 0.04$--$0.5$ pc range given a mass ratio, $q$, of the black holes, assuming the Monte Carlo average kick magnitude-dependence on $q$. We show various initial eccentricities assumed in the pre-kick disk, $e_0$. In all cases, the apsidal alignment is strongest when $q = 0.1$--$0.2$. For initially low to moderate eccentricity disks with $e_0 = 0$--$0.6$, the mean eccentricity converges at around 0.3 at a mass ratio $q \approx 0.06$. The $e = 0.3$ eccentric disk corresponds to our fiducial model and is denoted with a dotted gray line in both panels. From the top panel, this yields an estimate $q = 0.06 \pm 0.04$ corresponding to a companion mass estimate of $2^{+3}_{-1.2} \times 10^5 \ M_{\odot}$ where the intervals are of 3$\sigma$-confidence.

\begin{figure}[h!]
\centering
\includegraphics[width=\linewidth]{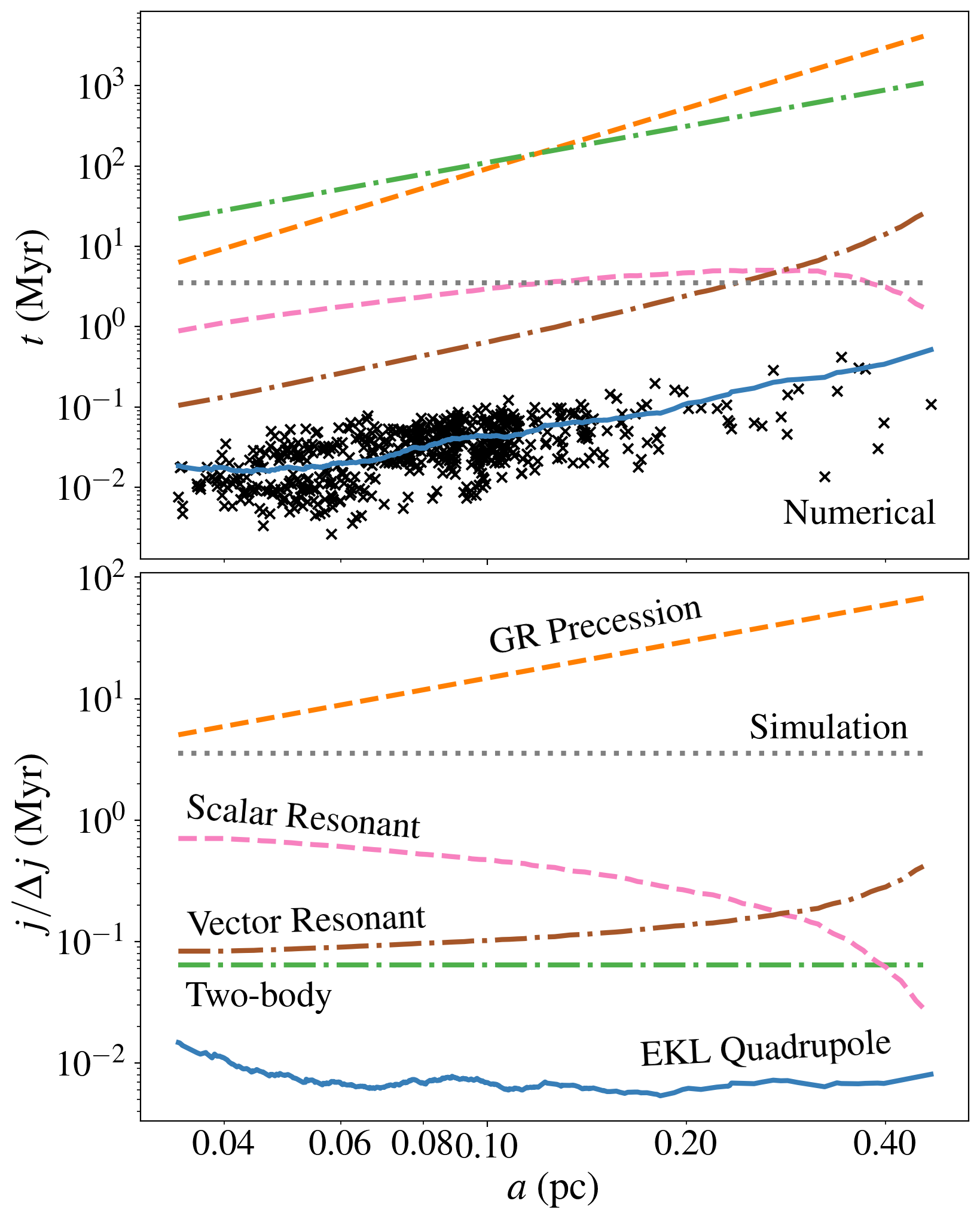}
\caption{(\textit{Top:}) Timescales of relevant physical processes: general relativistic precession, two-body relaxation, resonant relaxation, and the eccentric Kozai-Lidov mechanism (see \citealt{Naoz2016} for a review) for our simulations as a function of semi-major axis. The black ``x'' markers show the numerically estimated precession timescales. (\textit{Bottom:}) The corresponding relative changes to angular momentum for the same processes, an alternative comparison method suggested by \citealt{naoz2022} \citep[see also][]{melchor2024}. The linestyles are the same in both panels. The eccentric Kozai-Lidov quadrupole timescale matches the numerically estimated precession timescales quite well and is the most dominant driver of secular torques in our simulations. The general relativistic precession timescale is longer than the simulation time and is thus neglected in our current work. From the bottom panel, one can see that all relaxation processes are relevant within a simulation time.}
\label{fig:timescales}
\end{figure}

\section{Numerical Methods} \label{sec:numerical}

We compare timescales to determine which physical processes are important in our simulations (see Appendix \ref{app:timescales} for details). In the top panel of Figure \ref{fig:timescales}, we plot the relevant timescales as a function of the semi-major axis. In the bottom panel, we plot the corresponding relative angular momentum changes for the same processes, an alternative comparison scheme suggested by \citealt{naoz2022} \citep[see also][]{melchor2024}. The EKL quadrupole timescale \citep{naoz2013b, antognini2015} matches the numerically estimated precession timescales quite well and is by far the most dominant process in our simulations (see \citealt{Naoz2016} for a review). These timescales correspond to the precession of the argument of periapsis, $\omega$, by $2 \pi$. The general relativistic precession timescale adapted from \citet{naoz2013b} is longer than the simulation time at all semi-major axes and is thus neglected in our simulations. The various relaxation processes considered \citep{rauch1996, Hop06a} are all relevant within a simulation time, especially with respect to changes in angular momentum. A combination of these relaxation processes causes the efficient damping of apsidal alignment in our fiducial simulation (see Appendix \ref{app:ecc_vec}). It is important to note, however, that these relaxation processes are made arbitrarily shorter in our simulations due to the low $N$ of stars adopted in our current work since the relaxation timescale increases linearly with $N$ when conserving the stellar disk mass.

\begin{figure*}[h!]
\centering
\begin{tabular}{cccc}
$t=0$ Myr & $t = 2$ Myr \\
\includegraphics[width=0.4\linewidth]{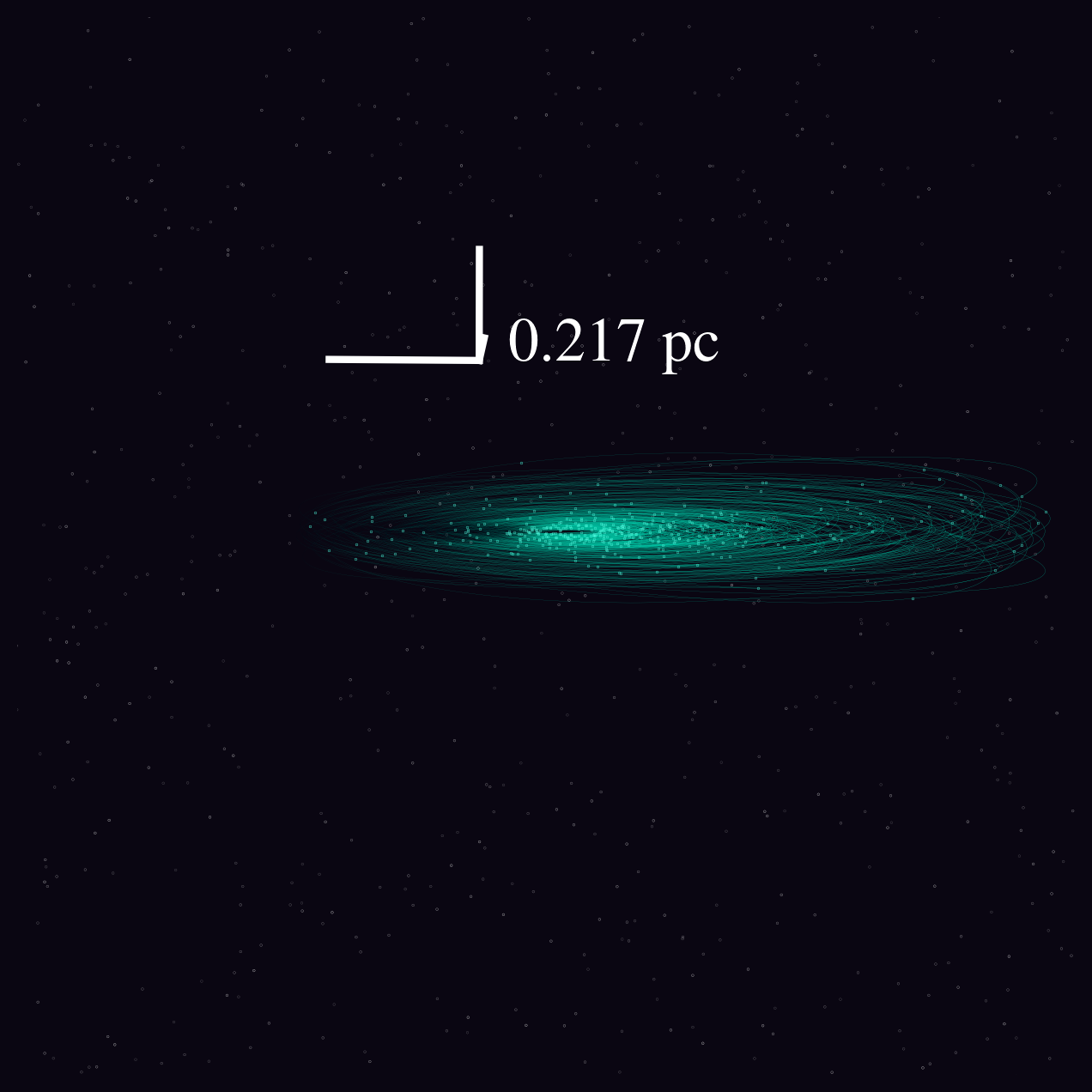} & \includegraphics[width=0.4\linewidth]{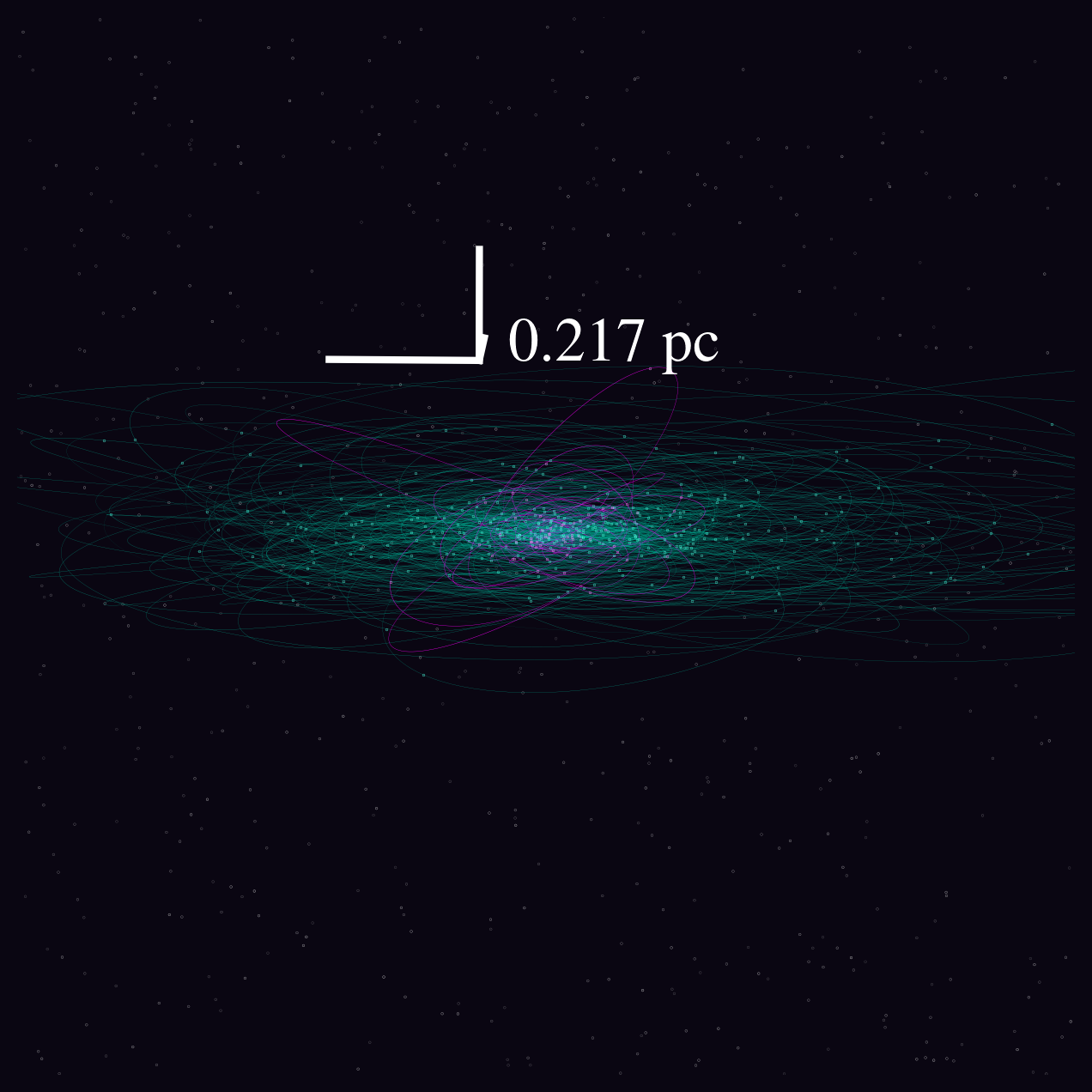} \\ 
\includegraphics[width=0.4\linewidth]{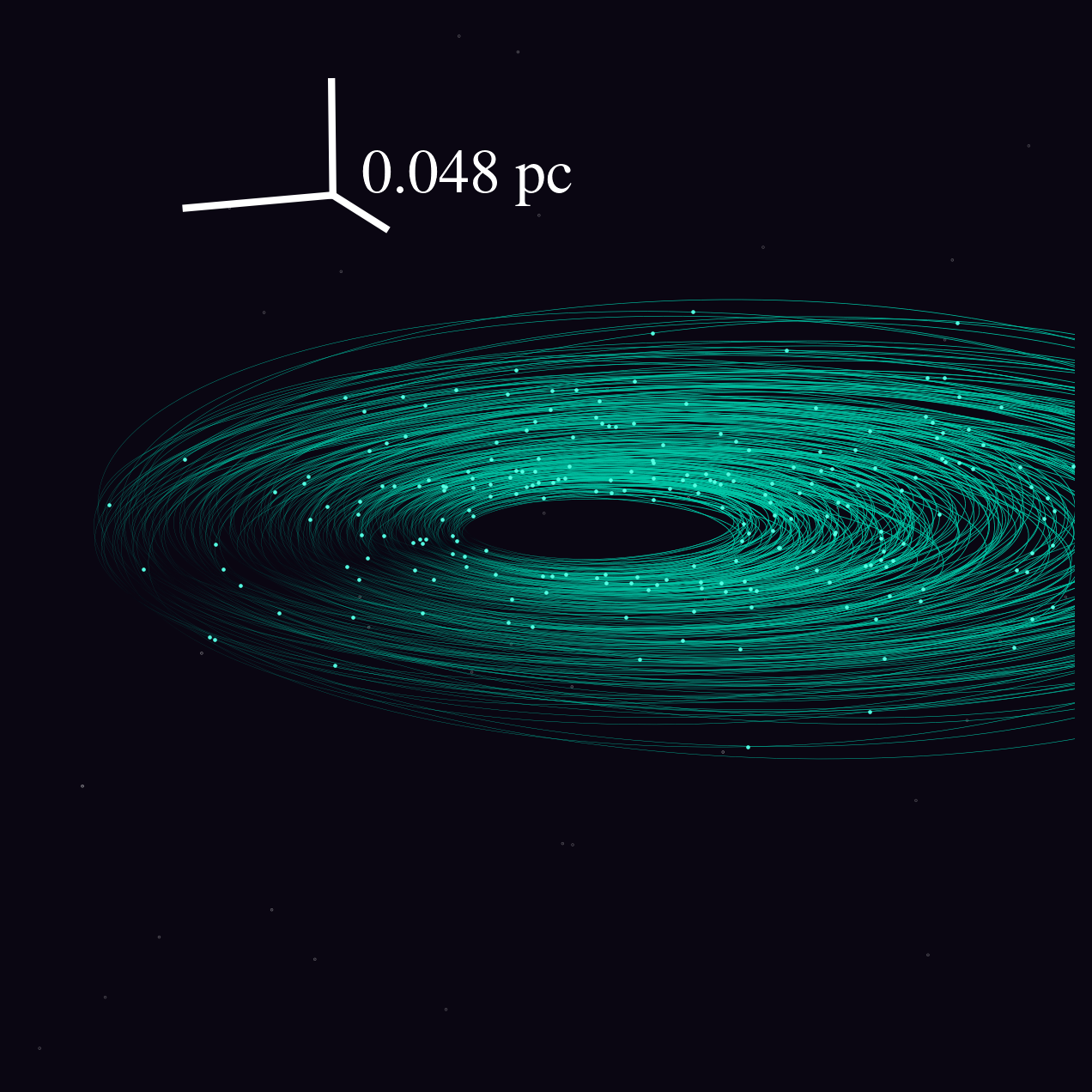} & \includegraphics[width=0.4\linewidth]{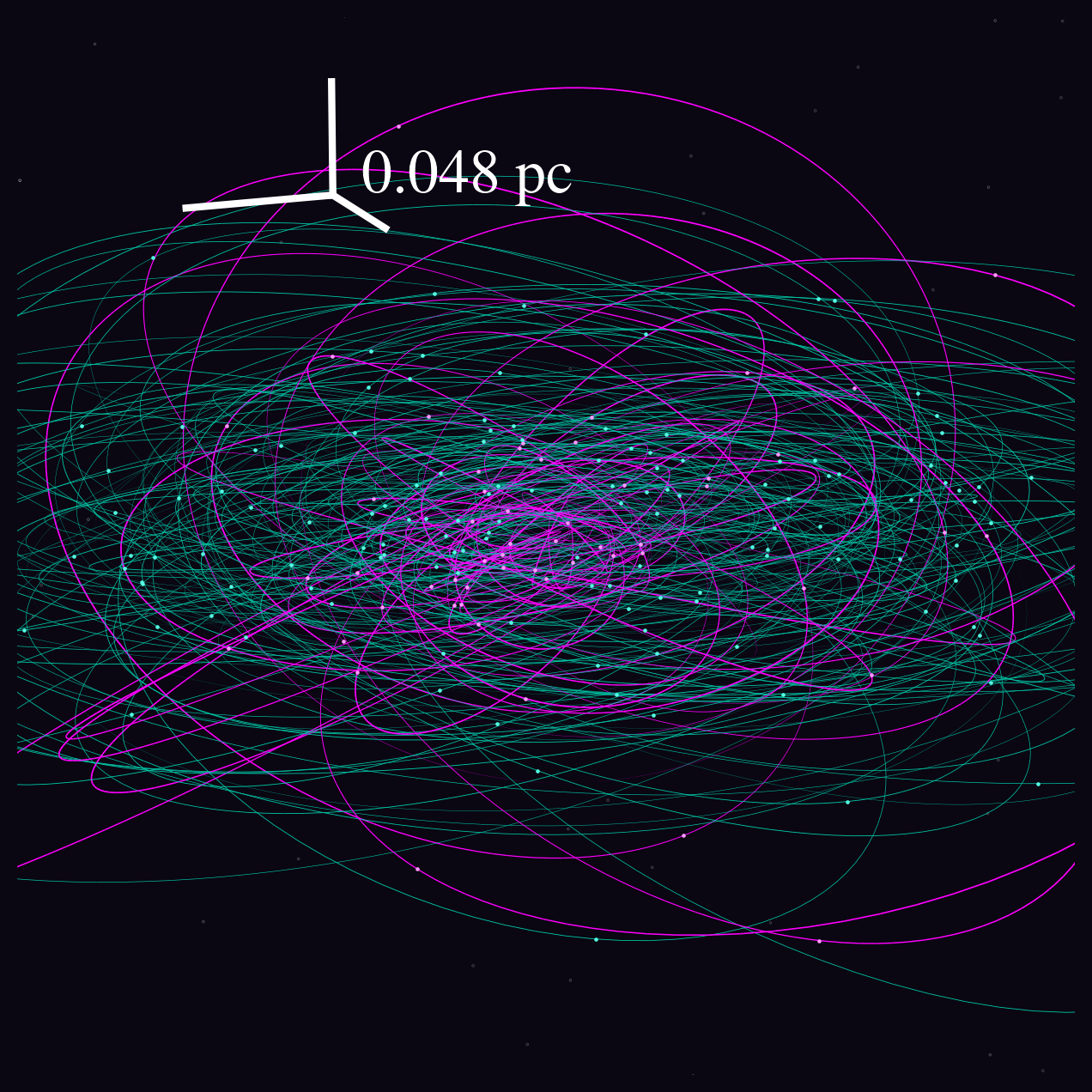} \\ 
\includegraphics[width=0.4\linewidth]{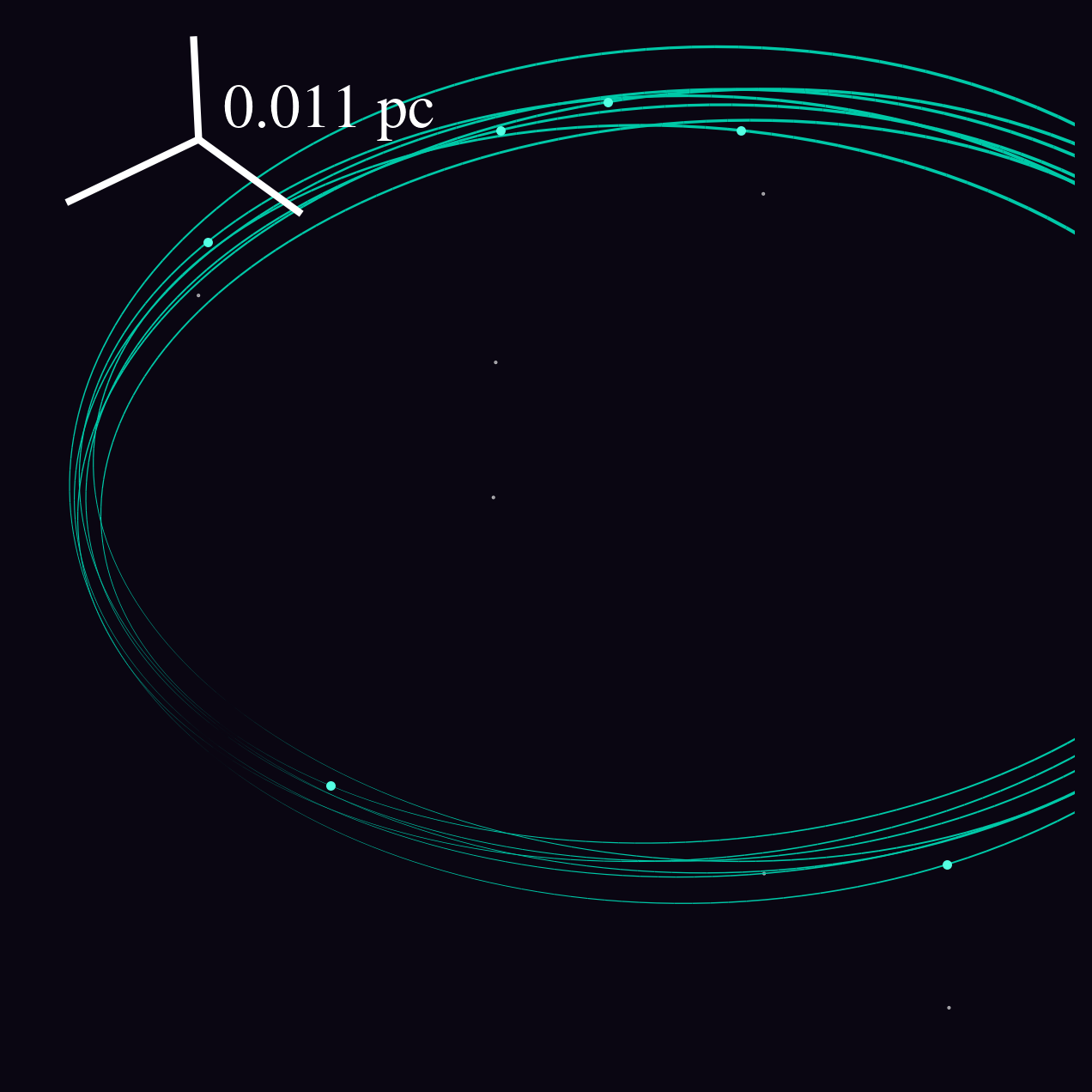} & \includegraphics[width=0.4\linewidth]{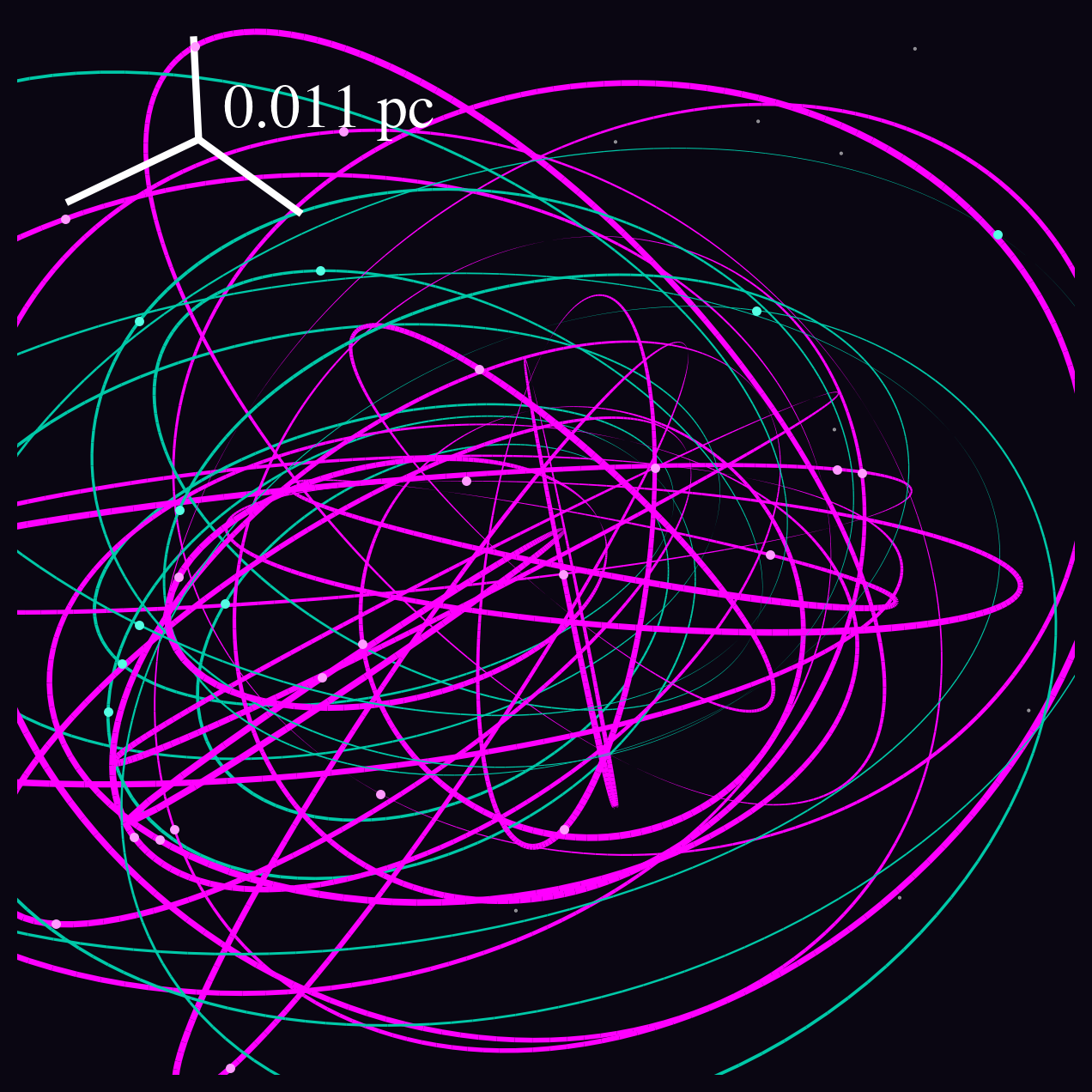} \\ 
\end{tabular}
\caption{Orbits zooming-in from $\approx 0.5$ pc to $\approx 0.02$ pc at (\textit{left:}) $t = 0$ and (\textit{right:}) $t=2$ Myr for our fiducial model simulation: $e = 0.3$, $\sigma_i = 3^\circ$. We emphasize orbits with $i>30^\circ$ in magenta, while low inclination orbits are plotted in teal green. After 2 Myr, the initially unoccupied region of $a < 0.04$ pc is populated with high-inclination, high-eccentricity orbits reminiscent of the S-stars. At length scales $> 0.1$ pc, orbits remain in a coherent disk with low inclinations while apsidal alignment is damped.}
\label{fig:orbit_3d}
\end{figure*}

\begin{figure*}[t!]
\centering
\includegraphics[width=\linewidth]{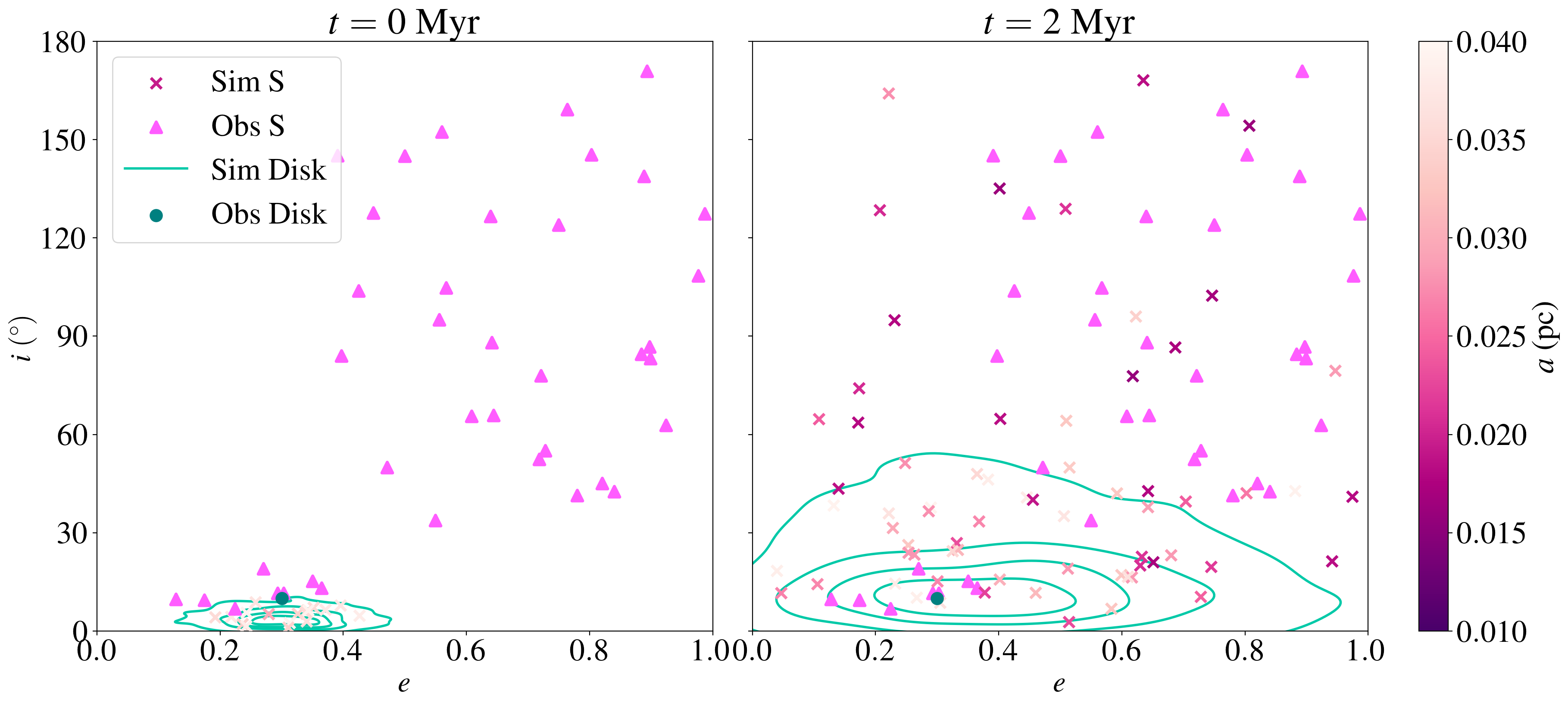}
\caption{The inclination and eccentricity distributions of the S-stars and disk stars at (\textit{left}:) $t = 0$ and (\textit{right}:) $t = 2$ Myr for our fiducial model. The simulation S stars are shown in magenta ``x'' markers, whereas the observed S star population taken from \citet{gillessen2017} is shown in pink triangle markers. The simulation disk stars are marked by teal-colored iso-proportional density contours which divide the population into equal quintiles, and the dark teal circle marker indicates the observed mean eccentricity and inclination of the disk stars \citep{yelda2014, lu2009, bartko2009}. The initial eccentric disk is thin and tightly clustered at $e = 0.3$ with no S-stars present. After 2 Myr, many stars have migrated inward of 0.04 pc and exhibit much more extended distributions in both eccentricity and inclination. The right panel shows a qualitative match between observations and simulations for both the S-stars and the disk stars.}
\label{fig:inc_ecc_dist}
\end{figure*}

According to the bottom panel of Figure \ref{fig:bh_kick}, one can reasonably expect an eccentric disk with mean $e = 0.3$ following a merger of mass ratio $q \approx 0.06$ if the mean pre-kick eccentricities are $e_0 \leq 0.6$. Under ideal conditions of a nearly circular disk and a mass ratio in the range $q = 0.1$--$0.2$, one can expect an eccentric disk with mean $e = 0.8$. We run a series of $N$-body simulations following the evolution of an eccentric disk with eccentricities $e = 0.3$ and $0.8$ to determine whether a Galactic Center-like structure within 0.5 pc can be induced in these scenarios.

We use the $N$-body package \texttt{REBOUND} with the \texttt{IAS15} integrator \citep{Rein2012, Rein2015}. In addition to comparing low and high mean eccentricities, we alter the disk thickness. The thin disk simulations have inclinations Rayleigh-distributed with scale parameter $\sigma_i = 3^\circ$ whereas the thick disk simulations have a scale parameter $\sigma_i = 30^\circ$. In each simulation run, we distribute $N = 800$ equal-mass particles in an eccentric disk of mass 4\% of the supermassive black hole mass, so $M = 4 \times 10^6 \ M_{\odot}$ and $M_{\rm{disk}} = 1.6 \times 10^5 \ M_{\odot}$. We adopt the clockwise disk stars' current semi-major axis range of $a = 0.04$--$0.5$ pc with a cusp-like density profile, $\Sigma \propto a^{-1}$. We set every particle to have an initial eccentricity value of $e = 0.3$ or $0.8$, and we restrict the longitude of periapsis, $\varpi$, to within a range of $3^\circ$ so that stellar orbits begin in apsidal alignment. Mean anomalies are uniformly distributed in $[0, 2\pi)$. Each simulation is integrated for approximately 3 Myr following the kick.

\section{Proof of Concept Numerical Example}
\label{sec:proof}

Our fiducial model is an eccentric disk with $e = 0.3$ and $\sigma_i = 3^\circ$ --- the low eccentricity, thin disk case. This setup reproduces many of the Galactic nuclear star cluster orbital properties in the central 0.5 pc within a few Myr. In Figure \ref{fig:orbit_3d}, we show the stellar orbits at various length scales, zooming in from $a \approx 0.5$ pc down to $a \approx 0.02$ pc for the fiducial model run comparing $t = 0$ and $t = 2$ Myr. After 2 Myr, the distributions of eccentricities and inclinations have settled into a steady state, remaining roughly the same until the end of the simulation. Orbits with $i > 30^\circ$ are emphasized in magenta. After 2 Myr, the initially unoccupied region within $0.04$ pc is populated with high inclination and high eccentricity S-stars due to a combination of relaxation and EKL-like secular torques. At longer length scales of $a > 0.1$ pc, the coherently-rotating disk is maintained throughout the simulation. The initial apsidal alignment is effectively erased during the few Myr evolution due to a combination of relaxation processes (see Appendix \ref{app:ecc_vec}).

In Figure \ref{fig:inc_ecc_dist}, we show the inclination and eccentricity distributions for the S-stars and disk stars during the same simulation at $t = 0$ and $t = 2$ Myr. We overplot the distributions from Galactic Center observations in both panels. After 2 Myr, the region within 0.04 pc is populated with S-stars, which exhibit much more extended distributions in both eccentricity and inclination. The simulation disk stars are consistent with an unimodal eccentricity distribution centered on $e = 0.3$ and a disk thickness of about $10^\circ$ \citep{yelda2014, lu2009, bartko2009}. The S-stars show a qualitative match between our simulation run and the observations from \citet{gillessen2017}, but the mean eccentricity and inclination of simulation S-stars are $e = 0.46$ and $i = 46^\circ$ which are notably lower than the observed S-stars with means $e = 0.61$ and $i = 79^\circ$, respectively, indicating a statistical discrepancy. We do not perform any formal statistical tests, however, since these results are derived from a single, proof-of-concept numerical experiment. A more detailed comparison between simulations and observations would require an extensive parameter study with more realistic models of nuclear star formation, black hole merger process, and eccentric disk formation.

\section{Conclusions and Implications for the Galactic Center}
\label{sec:discussion}

The S-stars at a distance $<0.04$ pc are on eccentric orbits that are isotropically distributed in inclination, in contrast with the surrounding clockwise disk at 0.05--0.5 pc that shows coherent rotation and only moderate eccentricities. The S-stars are challenging to explain since mechanisms for migration and excitation of eccentricities and inclinations need to work within 15 Myr, the estimated age of the young star cluster. In this paper, we explore the possibility that this morphology of young stars near Sgr A* is a dynamical imprint of a recent MBH merger in the Galactic Center.

We perform $N$-body simulations to show that many of the orbital properties of the clockwise disk and the S-star cluster in the central 0.5 pc of the Galactic Center can be explained from the dynamical evolution of an eccentric disk motivated by a gravitational recoil kick origin. The recoil kick is a consequence of a proposed MBH merger with Sgr A* in the recent past. The likely recoil kick magnitude based on the black hole mass ratio is calculated using a Monte Carlo approach from an analytical model \citep{lousto2010, lousto2012}, and the resulting apsidal alignment of the post-kick stellar disk is determined with a suite of \texttt{REBOUND} \citep{Rein2012} simulations. Finally, we conduct four $N$-body numerical experiments to explore how a Galactic Center-like structure can be reproduced as a natural dynamical consequence of a post-kick eccentric disk. Our main results are as follows:
\begin{enumerate}
    \item Our fiducial model starts with an apse-aligned, thin eccentric disk with $e = 0.3$ and Rayleigh-distributed inclinations with scale parameter $\sigma_i = 3^\circ$. In this scenario, the high eccentricity, high inclination S-star cluster, and the surrounding disk are induced within a few Myr due to secular, EKL-like torques. A schematic summary of the proposed mechanism can be found in Figure \ref{fig:schematic}.
    \item For low to moderate pre-kick eccentricities of $e_0 \leq 0.6$, an eccentric disk with $e = 0.3$ corresponds to a mass ratio of $q = 0.06 \pm 0.04$ or a companion mass of $2^{+3}_{-1.2} \times 10^5 \ M_{\odot}$. Given the young age of the S-star cluster, the merger would have had to happen in the last $\approx 10$ Myr.
\end{enumerate}

There are several important observational implications of the current work for the Galactic Center. As seen in Figure \ref{fig:bh_kick}, the mass ratio of the merger maps to a kick magnitude, which dictates the mean eccentricity of the lopsided disk, so this mechanism could potentially constrain the properties of the pre-merger black holes --- the progenitors of Sgr A*. Taking into account the pre-merger black hole spins and mass ratio, one can predict the final black hole mass, spin, and recoil kick from numerical relativity experiments \citep[e.g.,][]{Gonzalez2007b, Campanelli2007a, Campanelli2007b, Sperhake2020, Radia2021}. In particular, the final black hole spin and the recoil kick are correlated with each other. Recently, \citet{wang2024} suggested that a past massive black hole merger can produce the peculiar spin properties of Sgr A*. Using our proposed mechanism, one can better constrain the pre-merger black hole properties by combining constraints on the recoil kick magnitude with those on the current spin of Sgr A* \citep{eht2023, daly2024}.

The $e = 0.8$ eccentric disk scenario produces a much more extended, $\sim$pc central region made up of stars on high inclinations, extreme eccentricities, and strong apsidal alignment (see Appendix \ref{app:ecc_vec}). While it does not produce a Galactic Center-like structure, these results are applicable to nearby galaxies that host an asymmetric nucleus like the Andromeda galaxy \citep{Tremaine1995}. We emphasize that every MBH merger has an associated recoil kick --- many of the puzzling nuclear features in nearby galaxies and the Milky Way may be explained by the dynamical evolution of recoil kick-induced eccentric disks starting with varying degrees of alignment and lopsidedness. Since the post-kick eccentric disk depends on the kick magnitude, which in turn is dictated by the mass ratio and spins of the pre-merger black holes, the proposed mechanism offers a unique avenue to explore the merger history of the Milky Way Galactic Center and other nearby galactic nuclei.

\section{Acknowledgements} \label{sec:acknowledgements}

S.N. acknowledges the partial support from NASA ATP 80NSSC20K0505 and the NSF-AST 2206428 grant and thanks Howard and Astrid Preston for their generous support.
AM gratefully acknowledges support from the David and Lucile Packard Foundation. This work utilized resources from the University of Colorado Boulder Research Computing Group, which is supported by the National Science Foundation (awards ACI-1532235 and ACI-1532236), the University of Colorado Boulder, and Colorado State University.

\appendix

\renewcommand\thefigure{\thesection\arabic{figure}}
\setcounter{figure}{0}

\section{Analytics for the Gravitational Recoil Kick}
\label{app:kick}

To obtain the recoil kick, we take the analytical model from \citet{lousto2010, lousto2012} as outlined in \citet{fragione2023}. Their prescription suggests the kick vector
\begin{equation} \vec{v}_{\rm{kick}} = v_m \ \hat{x} + v_{\perp} ( \cos \xi \ \hat{x} + \sin \xi \ \hat{y} ) + v_{\parallel} \ \hat{z} \ ,
\end{equation}
where 
\begin{equation}
v_m = A \eta^2 \sqrt{1 - 4 \eta} (1 + B \eta) \ , 
\end{equation}
\begin{equation}
v_{\perp} = \frac{H \eta^2}{1 + q} (\chi_{2, \parallel} - q \chi_{1, \parallel}) \ , 
\end{equation}
and
\begin{equation}
\begin{aligned} v_{\parallel} &= \frac{16 \eta^2}{1 + q} [V_{1,1} + V_A \tilde{S}_{\parallel} + V_B \tilde{S}_{\parallel}^2 + V_C \tilde{S}_{\parallel}^3] \\
& \times |\chi_{2, \perp} - q \chi_{1, \perp}| \cos (\phi_\Delta - \phi_1) \ . 
\end{aligned}
\end{equation}
given a mass ratio $q = m_1/M_2$, a corresponding asymmetric mass ratio $\eta = q/(1+q)^2$, and dimensionless spin vectors $\vec{\chi}_{1}$ and $\vec{\chi}_{2}$. The subscripts 1 and 2 refer to the less massive and more massive black hole, and the subscripts $\perp$ and $\parallel$ refer to components perpendicular and parallel to the orbital angular momentum axis, respectively. The vectors $\hat{x}$ and $\hat{y}$ are orthogonal unit vectors in the orbital plane. The vector $\tilde{S}$ is defined as
\begin{equation}
\tilde{S} = 2 \ \frac{\vec{\chi}_{2} + q^2 \vec{\chi}_{1}}{(1+q)^2} \ ,
\end{equation}
$\phi_1$ is the phase angle of the binary, and $\phi_\Delta$ is the angle between the in-plane component of
\begin{equation}
\vec{\Delta} = M^2 \ \frac{\vec{\chi}_{2} - q \vec{\chi}_{1}}{1+q} \ ,
\end{equation}
and the infall direction at merger. For our work, we assume that the angle $(\phi_\Delta - \phi_1)$ is uniformly distributed in $[0, 2 \pi)$. The values for the constants we adopt are $A = 1.2 \times 10^4 \ \rm{km} \ \rm{s}^{-1}$, $H = 6.9 \times 10^3 \ \rm{km} \ \rm{s}^{-1}$, $B = -0.93$, $\xi = 145^\circ$, $V_{1,1} = 3678 \ \rm{km} \ \rm{s}^{-1}$, $V_A = 2481 \ \rm{km} \ \rm{s}^{-1}$, $V_B = 1793 \ \rm{km} \ \rm{s}^{-1}$, and $V_C = 1507 \ \rm{km} \ \rm{s}^{-1}$ \citep{Gonzalez2007a, Lousto2008, lousto2012}. This model is in good agreement with full numerical relativity results even in the intermediate mass ratio regime of $q \sim 0.1$ \citep{gonzalez2009}.

\section{Modeled Physical Processes and Timescales}
\label{app:timescales}

The eccentric Kozai-Lidov quadrupole timescale is calculated as
\begin{equation}
t_{\rm{quad}} \sim \frac{a_{\rm{out}}^3 (1 - e_{\rm{out}}^2)^{3/2} \sqrt{M}}{a^{3/2} M_{\rm{out}} \sqrt{G}} \ ,
\label{eqn:kozai}
\end{equation}
where for a given star at a semi-major axis $a$, we compute $a_{\rm{out}}$, $e_{\rm{out}}$, and $M_{\rm{out}}$ as the mean semi-major axis, mean eccentricity, and total mass of neighboring stars outward of $a$, and $M$ is the mass of the supermassive black hole \citep{naoz2013b, antognini2015}. In our analysis, we consider neighboring stars in the range $a$ to $2a$, so that $a_{\rm{out}} \propto a$. The general relativistic precession timescale we use is the first-order post-Newtonian timescale estimate of
\begin{equation}
t_{\rm{1PN}} \sim \frac{2 \pi a^{5/2} c^2 (1-e^2)}{3 G^{3/2} M^{3/2}} \ ,
\label{eqn:gr}
\end{equation}
where $a$ and $e$ are the semi-major axis and eccentricity of the star and $M$ is the mass of the supermassive black hole \citep{naoz2013b}. For both Equations \ref{eqn:kozai} and \ref{eqn:gr}, we explicitly assume that $M \gg M_{\rm{disk}}$, the mass of the stellar disk. We also include the relevant timescales for relaxation processes. The two-body relaxation, scalar resonant relaxation, and vector resonant relaxation timescales are computed as
\begin{equation} t_{\rm{2body}} = \frac{M^2}{{m_*}^2 N \ln{N}} \ t_{\rm{orb}} \ ,
\end{equation}
\begin{equation} t_{\rm{scalar}} = \frac{m_* N_{\rm{out}}}{M} \ t_{\rm{2body}} \ ,
\end{equation}
and
\begin{equation}
t_{\rm{vector}} = \frac{M}{2m_* \sqrt{N_{\rm{out}}}} \ t_{\rm{orb}} \ ,
\end{equation}
respectively, where $M$ is the mass of the black hole, $m_*$ is the mass of a star, $t_{\rm{orb}}$ is the orbital period, $N$ is the number of stars, and $N_{\rm{out}}$ is the number of stars outward of the semi-major axis considered \citep{rauch1996, Hop06a}. In the top panel of Figure \ref{fig:timescales}, we plot these timescales directly. While resonant relaxation processes are relevant within a simulation time, the precession is predominantly governed by the EKL quadrupole timescale. In the bottom panel, we plot the timescales as they pertain to changes in angular momentum. The notable change is due to the fact that $j/\Delta j \propto \sqrt{t}$ for relaxation processes and the EKL quadrupole timescale, but $j/\Delta j \propto t$ for the general relativistic precession timescale. For relative angular momentum changes, all relaxation processes play a significant role within a simulation time.

\section{Apsidal Alignment and Comparisons between Numerical Experiments}
\label{app:ecc_vec}

\begin{figure*}[t!]
\centering
\includegraphics[width=\linewidth]{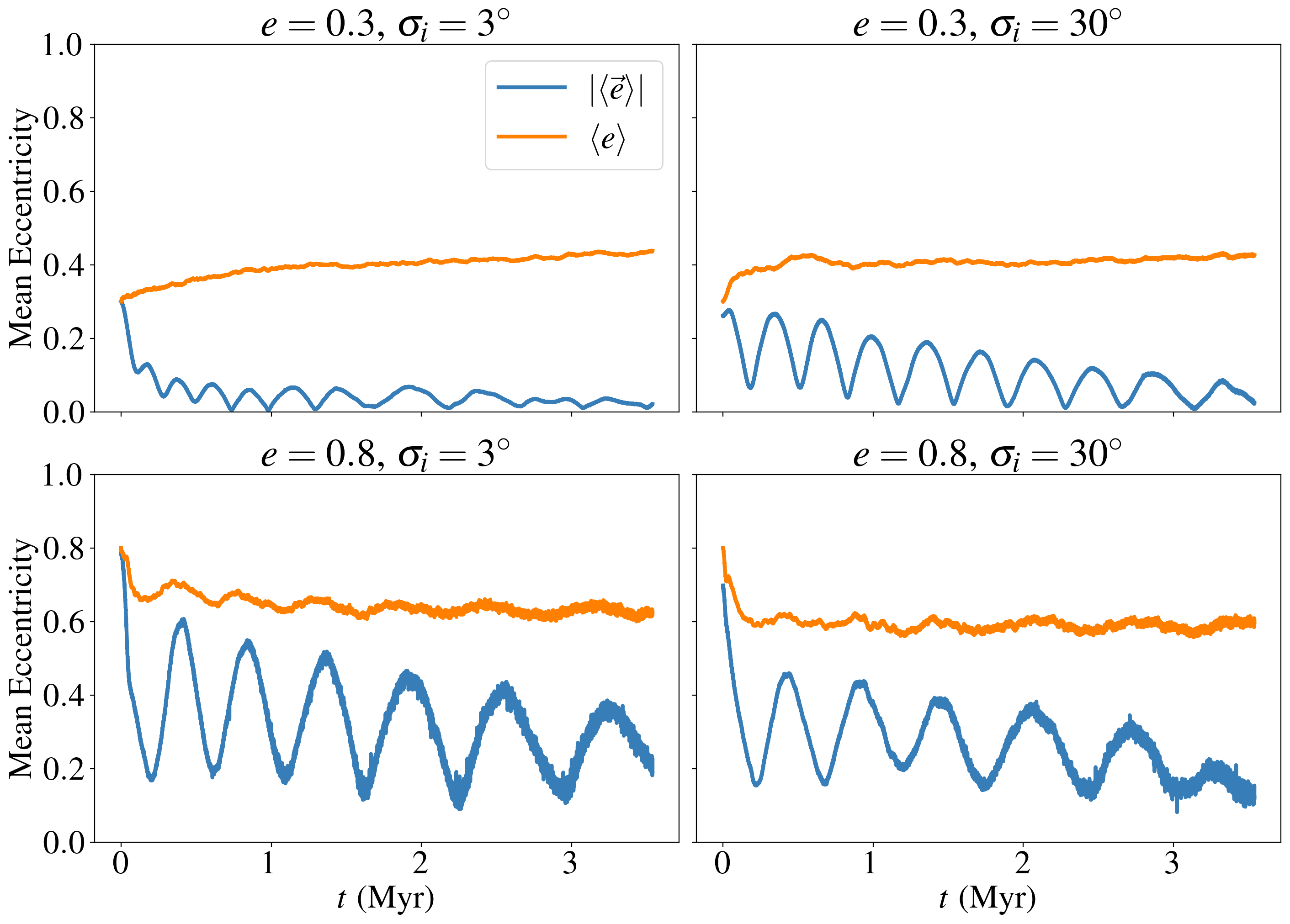}
\caption{Evolution of apsidal alignment (blue) and the mean scalar eccentricity (orange) for our simulation runs. The scalar eccentricity only takes into account the magnitude of each star's orbit, whereas $| \langle \vec{e} \rangle| $ takes into account both the eccentricity magnitude as well as alignment. For $e = 0.3$ simulations, the apsidal alignment is erased during the course of the simulation. The alignment is maintained and the eccentric disk is stable for the $e = 0.8$ simulation runs.}
\label{fig:mean_ecc}
\end{figure*}

In Figure \ref{fig:mean_ecc} in the top left panel, we show the evolution of the eccentricity alignment over time for our fiducial model. The blue line which plots $| \langle \vec{e} \rangle |$ shows that the alignment is erased within a few Myr, while the orange line plotting the mean scalar eccentricity grows over time. The fiducial model of a weakly aligned, thin eccentric disk is able to reproduce the eccentricity and inclination distributions of the S-star cluster and surrounding disk while losing apsidal alignment within a few Myr.

We compare the fiducial model to the other numerical experiments we performed. We note the following key differences:

\begin{itemize}
    \item The semi-major axis range of the S-stars is much smaller and closer to the observed $\leq$0.04 pc range in the $e = 0.3$ scenario rather than the $e = 0.8$ case. While the $e = 0.3$ simulation only induces high inclinations in the central region $< 0.04$ pc, the $e = 0.8$ case has an extended population of high inclination orbits out to $\sim$pc scales.
    \item From Figure \ref{fig:mean_ecc} in the bottom row, we see that apsidal alignment cannot be completely erased for eccentric disks with $e = 0.8$. This is due to the stability mechanism of eccentric disks \citep{Madigan2018}. The damping of the mean eccentricity vector is also slower for a thicker disk because the relaxation timescale is longer. The mean scalar eccentricity increases over time for $e = 0.3$ but decreases over time for $e = 0.8$.
    \item For the thick disk cases, the distinction between the nearly isotropic S-star cluster and the coherently-rotating disk is less pronounced. The nearly isotropic configuration extends to larger distances and the inclination distribution does not vary significantly as a function of semi-major axis.
    \item The mean scalar eccentricity changes gradually over time, but the distribution to first-order retains the initial mean eccentricity and spreads diffusively. Thus, the resulting eccentricity distribution starting with $e = 0.3$ evolves into a unimodal distribution centered on $e = 0.3$ consistent with the observed disk star population. The $e = 0.8$ simulations, on the other hand, have a mean scalar eccentricity that is significantly higher than the observed mean throughout the simulation.
\end{itemize}

\bibliographystyle{aasjournal}
\bibliography{ms}

\end{document}